\documentstyle[float,twocolumn,epsfig,prl,aps]{revtex}

\begin{document}
\newfloat{figure}{ht}{aux}

\draft
\twocolumn[\hsize\textwidth\columnwidth\hsize\csname@twocolumnfalse\endcsname

\title{Non-trivial fixed point structure of the two-dimensional
$\pm$J 3-state Potts ferromagnet/spin glass}

\author{Erik S.\ S\o rensen$^1$, Michel J.\ P.\ Gingras$^2$ and David~A.~Huse$^3$}
\address{$^1$Laboratoire de Physique Quantique, IRSAMC, Universit\'e Paul
Sabatier, F-31062, Toulouse, Cedex 4, France\\
$^2$Department~of~Physics, University of Waterloo, Waterloo, Ontario, N2L-3G1
Canada\\
$^3$Physics Department, Princeton University, Princeton, NJ 08544, USA}

\date{\today}

\maketitle

\begin{abstract}
The fixed point structure 
of the 2D 3-state random-bond Potts model with a bimodal ($\pm$J)
distribution of couplings 
is 
for the first time 
fully determined
using numerical renormalization group techniques.
Apart from the pure and $T=0$ critical fixed points, {\it two} other non-trivial fixed
points are found.  One is the critical fixed point for the random-bond, but
unfrustrated, ferromagnet. The other is a bicritical
fixed point analogous to the bicritical Nishimori fixed point found in
the random-bond frustrated Ising model.   
Estimates of the associated critical exponents are
given for the various fixed points of the random-bond Potts model.
\end{abstract}
\vskip2pc]

%============================================================================

% BODY OF PAPER

Over the past decades the study of random systems has attracted considerable
attention. In particular the possibility of a spin-glass phase has been
studied in great detail, as a result of which 
it now seems well established that no finite temperature equilibrium
phase transition to a spin-glass phase occurs in two dimensions~\cite{SGs}. 
However, even in the absence of a glassy phase a rich fixed point structure
can occur. In particular, the transition to 
an ordered phase in the presence of disorder can be controlled by new 
random fixed points. Recently, it has been suggested that while pure systems
can display critical behavior in many different universality classes,
the number of universality classes may be more limited when
disorder is present~\cite{univ}. In addition, current progress in the
understanding of quantum phase transitions in disordered systems as
well as other complex systems has shown that many of these models are in
the same universality class as certain disordered classical statistical
mechanical models~\cite{LFRG}. It is therefore highly desirable to
undertake a careful
investigation of the complete fixed point 
structure of some key models, determining
all stable and {\it unstable} fixed points.
While considerable progress towards
the understanding of random fixed points has been 
made using analytical techniques~\cite{ludwig,shankar,dotsenko,luther} 
and series expansions~\cite{singh}, it would seem 
that a first step towards such a
classification
of the fixed point structure would have to come from numerical
work that does not rely on perturbative methods and where frustration can
be taken into account satisfactorily. McMillan~\cite{McMill}
investigated the fixed point structure of a two-dimensional, frustrated 
random-bond Ising model using the domain wall 
renormalization group (DWRG) method. In this
case a bicritical point occurs, separating the ferromagnetic-to-disordered
critical phase boundary into two parts.  The part at higher temperature
and weaker disorder is governed by the pure critical fixed point, 
while the lower-temperature portion of the phase
boundary is governed by the zero-temperature
fixed point that also governs the ferromagnetic-to-spin glass 
transition at zero temperature.  For the random Ising model this
bicritical point occurs on the
so-called Nishimori
line~\cite{nishimori}, where there is a special gauge symmetry,
and the ferromagnetic and spin-glass correlations and susceptibilities
are identical~\cite{singh}.

In the
present paper we apply a variant of the renormalization group (RG) method of
McMillan~\cite{McMill}
to the 3-state Potts model and determine all the fixed points. For this model
it is known from the Harris criterion~\cite{harris} 
that the pure fixed point is unstable against disorder and the simplest
possible scenario for the RG flow would be a flow out 
from the pure fixed point that goes to
the zero-temperature fixed point separating the
ferromagnetic and spin-glass phases. However, we show that this is {\it not}
the case and that {\it two} additional 
non-trivial random fixed points occur at nonzero disorder and nonzero
temperature. 
In agreement with analytical work using only ferromagnetic 
(unfrustrated) disorder~\cite{ludwig,dotsenko}, an unfrustrated
random critical fixed point is found.  However,
in addition to this critical fixed point 
a bicritical fixed point {\it also} occurs in this model, 
even though the model does {\it not} have the gauge symmetry that is
used to define the Nishimori line in the Ising model.
We believe that this fixed point structure is of a rather general nature, in
the sense that it should be similar for
many other models for which disorder is relevant.

The Hamiltonian we use is:
\begin{equation}
H=-\sum_{<i,j>}
J_{ij}h(n_i,n_j),\ \ n_i=1\ldots q,
\label{hamiltonian}
\end{equation}
with $h(n_i,n_j)=\cos(2\pi(n_i-n_j)/{q})$,
and $q=2$ (Ising) or $q=3$ (Potts).
The sum is over all nearest-neighbor pairs of sites on
a square lattice.
We consider the situation where the bonds $J_{ij}$ in 
Eq.~\ref{hamiltonian} are distributed randomly, and
given by a quenched biased bimodal probability
distribution:
${\cal P}(J) \; = \; x\delta (J - 1) + (1-x)\delta (J + 1)  . 
$
As an application of this type of model,
it has been suggested that the orientational 
freezing in molecular glasses, such as 
N$_2$-Ar and KBr-KCN, can be partially described by a three-dimensional 3-state 
Potts spin glass~\cite{Binder_Reger}.

The way we apply the DWRG
is slightly novel and we therefore
begin by reviewing
the method and the scaling ideas behind our approach.
The DWRG as
proposed by McMillan~\cite{McMill} estimates
the singular part of free energy by 
calculating the domain wall energy, $[\Delta F]$, 
and its standard deviation, $\sigma(\Delta F)$,
for systems of size $L\times L$. 
Here $[\cdot]$
denotes disorder averaging.
We calculate the total
free energy difference
between periodic (p) and twisted (t) boundary conditions,
$\delta F_m = {F}_{\rm t}-{ F}_{\rm p}$, 
for a long cylinder of size $L\times M$ where $M=Lm$,
and $m$ is
a large integer running up to of order $10^5-10^6$. 
The proper generalization of twisted boundary conditions
to $q>2$ is to change the local Hamiltonian across the boundary
in the $L$-direction
as follows~\cite{denNijs}: 
$h(n_1,n_L) = \cos(2\pi((n_1-n_L+1)/{q})$. This change is done at a seam going 
{\it along} the entire length of the cylinder. 
$\Delta { F}$ for each $L \times L$ subsystem is
simply defined as $\Delta F = \delta F_m - \delta F_{m-1}$.
The free energies are evaluated exactly using standard transfer matrix
techniques~\cite{TM-refs} and the only errors in the calculation therefore
stem from the incomplete disorder averaging.

{\it Pure systems.}
In the absence of disorder
it is known from hyperscaling that the singular
part of the free energy density 
scales as the inverse of a correlation volume
\begin{equation}
\frac{f_s}{k_bT_c} = \frac{C}{\xi^d}.
\label{eq:hscaling}
\end{equation}
Since $f_s \sim \Delta { F}/L^d$
it follows that an appropriate finite-size scaling form for $\Delta F$ is
\begin{equation}
\frac{\Delta { F}}{k_bT} = A g(\delta L^{1/\nu}),
\label{eq:scalepure}
\end{equation}
with $g$ a universal function and $\delta=|T-T_c|$.
Hence, the critical point, $T_c$, can be located by 
standard methods, i.e. by tracing
$\Delta F$ as a function of $T$ for several different system sizes
and locating the point where the lines cross.
Here $A$ is the {\it universal} amplitude for the spin stiffness of the system.
The universality of
$A$ has been investigated extensively at finite temperature
transitions~\cite{denNijs,Night,Cardy,Abraham,Mon}, 
and is known exactly
for the {\it pure} $q=2,3,4$ Potts models as a function of 
the aspect ratio $s=1/m$~\cite{denNijs}.
In the limit $s\to 0$ it follows from conformal invariance
that 
$A=\pi \eta$~\cite{Cardy}, where $\eta$ is the magnetic
critical exponent.
Since we use
an aspect ratio of $s=1/m$, essentially zero, we see that
$\frac{\Delta F}{k_bT_c}= \pi\eta.$
Hence, $\Delta F$ directly measures a {\it bulk} 
critical exponent.

{\it Disordered systems.}
Just as $\Delta F/k_BT$ is a {\it universal} amplitude at the
pure fixed point it is natural to assume that any random fixed point occurring
at {\it finite, nonzero} temperature and disorder
will be characterized by a universal
{\it distribution} of $\Delta F/k_BT$. 
In particular, the mean $[\Delta F]/k_BT$ and
the standard deviation $\sigma(\Delta F)/k_BT$ 
are then universal~\cite{FisherHuse,BrayMoore}.
The generalization of the finite
size scaling relation Eq.~(\ref{eq:scalepure}) to include
disorder is therefore:
\begin{equation}
\frac{[\Delta { F}]}{k_bT} = g_X(\delta L^{1/\nu},\epsilon L^{\lambda_2}),
\label{eq:scaled1}
\end{equation}
valid close to any finite temperature fixed point $X$
at $(x^*,T^*)$, with $g_X$
a universal scaling function that is, however, different for
each distinct fixed point $X$. $\sigma(\Delta F)/k_BT$ 
obeys a similar finite-size scaling form.
Here $\delta$ is a linear combination of 
$\Delta x = |x-x^*|, \Delta T=|T-T^*|$,
that is the eigenvector of the RG flow that moves
one away from the phase boundary;
$1/\nu$ is the corresponding leading eigenvalue.
$\epsilon$ is another linear combination that is 
the next subrelevant ($\lambda_2 > 0$)
or irrelevant ($\lambda_2 < 0$) eigendirection 
of the RG flow that is tangent to the
phase boundary.  Thus we can estimate the fixed points
$(x^\star,T^\star)$ as the points in the phase diagram where
$([\Delta F],\sigma(\Delta F))$ are independent of $L$.
We obtain a picture of the RG flow by measuring $[\Delta F]$
and $\sigma(\Delta F)$ at various $(x,T)$ points and seeing
how they vary as $L$ is increased at constant $x$ and $T$.

{\it Zero temperature.} If the transition is controlled by a zero temperature
fixed point $Z$, then hyperscaling, Eq.~(\ref{eq:hscaling}), is {\it not}
valid, and we cannot use
Eq.~(\ref{eq:scalepure}) as our starting point. 
The appropriate finite-size scaling form is instead
\begin{equation}
[\Delta { F}]= L^{\theta}g_Z(\delta L^{1/\nu},\epsilon L^{-\theta})  ,
\label{eq:scalezero}
\end{equation}
with $\theta > 0$.
Hence, $[\Delta { F}]$ is {\it not} universal along the part of the phase
boundary where the flow is controlled by a zero-temperature fixed point.
Instead, $[\Delta F]$ grows with 
increasing $L$ as $L^{\theta}$ at the critical
point. 
At small temperatures we expect that $\delta \sim |x-x_c|$ and
$T \sim \epsilon$, i.e. that the disorder $x$ is the relevant flow direction.
A positive $\theta$ implies that the flow is
{\it into} the zero-temperature fixed point and hence that temperature is
an irrelevant variable.

We now proceed as follows:
In order to accommodate both finite temperature and
zero temperature fixed points in the same plot we consider the two quantities
$r=\sigma(\Delta F)/[\Delta F]$ and $f=k_bT/[\Delta F]$. Any fixed point
should be characterized by universal values of these quantities 
$(r^\star,f^\star)$. In particular we know that for the pure fixed point we
have $(r^\star,f^\star)= (0,(\pi\eta)^{-1})$.
A ferromagnetic phase will be
characterized by $(r,f) \to (0,0)$ whereas in 
the paramagnetic phase $[\Delta F] \to 0$.
we can now directly determine the
complete fixed point structure by calculating 
$(r(L),f(L))$ 
for different values of $L$ on a grid of $(x,T)$ close to the
phase boundary. By connecting points, $(r(L),f(L))$ 
corresponding to $L$ and $L'$ calculated at the same $(x,T)$,
with an arrow terminating at the larger
of $L, L'$, the flow becomes clearly visible and ``fixed points"
$(r^\star,f^\star)$ appear where {\it both} $f$
and $r$ are independent of $L$. 
The linear flow around a fixed point should be characterized by
\begin{equation}
\left(\matrix{ f(L^\prime)-f^\star \cr 
               r(L^\prime)-r^\star\cr }\right)
	       = M
\left(\matrix{ f(L)-f^\star\cr 
               r(L)-r^\star\cr }\right).
\end{equation}
Through a least-square fit to this equation for a number of points close
to the fixed point the matrix $M$ as well as $(r^\star,f^\star)$ can
be determined and consequently the eigenvalues, $\lambda_1=1/\nu,\lambda_2$,
and eigenvectors of the linear
flow can be determined.
The
magnetic exponent $\eta$ is estimated by exact calculations
of $[<m^2>]\sim L^{2-\eta}$, the square of the magnetization per site, directly at
the fixed point.

\begin{figure}
  \begin{center}
  \epsfig{file=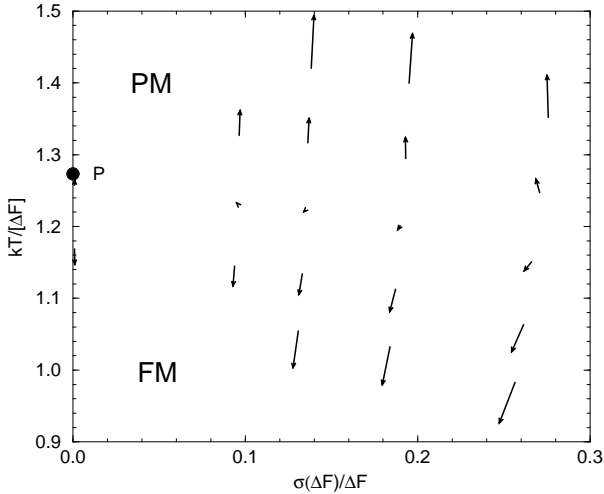,width=8.0cm}
  \caption{The flow close to the pure fixed point for the 2D Ising model
  with bimodal ferromagnetic disorder 
 ${\cal P}(J_{ij}) = (1-x)\delta(J_{ij}-1) + x\delta(J_{ij}-0.3)$.
 $P$ is the exactly known pure fixed point.
 In most cases 1 million $L \times L$ blocks were measured. 
   }\label{fig:q2jp0.3}
  \end{center}
\end{figure}
In order to establish that our 
approach is sensitive even to {\it marginal } flow
we first consider the two-dimensional Ising model with {\it
ferromagnetic} disorder. In this case it seems well
established~\cite{dotsenko,ludwig,shankar} that this type of disorder
is {\it marginally irrelevant} at the pure 
critical point. Hence the flow should be towards the 
pure critical fixed point $P$. Our results for this case are 
shown in Fig.~\ref{fig:q2jp0.3} for a bimodal distribution of couplings
of strength $J$ and $0.3J$. 
The arrows connect results for $L=6$ and $L'=8$.
It is clear that the flow in this
case indeed is {\it towards} $P$, located at $(0,4/\pi)$, 
in agreement with the theoretical studies~\cite{dotsenko,ludwig,shankar}.

\begin{figure}
  \begin{center}
  \epsfig{file= 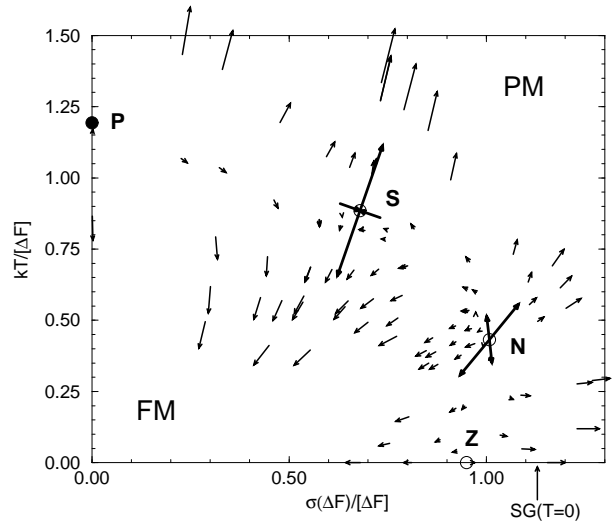,width=8.0cm}
  \caption{The flow diagram for the q=3 
Potts model. 
  $P$ is the exactly known pure critical fixed point, 
  $S$ the random critical fixed point,
  $N$ the analogue of the Nishimori bicritical fixed point, and $S$ is
  the zero temperature fixed point.
  The bold arrows indicate the numerically determined RG
  eigenvectors. The {\bf PM} and {\bf FM} indicate the paramagnetic and
  ferromagnetic phase. In most cases 200,000 
  $L \times L$ blocks were measured.
  }\label{fig:q3diag}
  \end{center}
\end{figure}
We now turn to a discussion of our main results for the 3-state Potts model.
The flow diagram is shown in Fig.~\ref{fig:q3diag} with $L=6$ and
$L'=8$. Rather clearly, four
different fixed points are visible along the phase boundary of
the ferromagnetic phase. They are: the pure critical fixed point $P$, the
random critical fixed point $S$, the bicritical fixed point $N$ and the 
zero-temperature fixed point $Z$. Close to the latter fixed point
our calculations have been performed exactly at zero temperature.
The RG flow along the phase boundary is away from $P$ and $N$ and
towards $S$ and $Z$. Also indicated (bold arrows) in Fig.~\ref{fig:q3diag}
are the numerically determined RG eigenvectors at the fixed points $S$ and
$N$.
At the {\it stable fixed point} ($S$), with approximately
$(x^\star,T^\star)\sim(0.93,1.16)$ 
$(r^\star,f^\star)=(0.68(9),0.89(3))$
we find $\lambda_1=0.94-1.04,\ \lambda_2\sim-.2$, and hence
$\nu=1/\lambda_1\sim 0.96-1.06$. Using $[<m^2>]$ we find $\eta=0.23-0.28$. Both
of these exponents are in good agreement with analytical
results~\cite{ludwig,dotsenko} as well as Monte Carlo work~\cite{Pico},
for the $q=3$ Potts model with ferromagnetic disorder,
so $S$ appears to be the same random fixed point
as in the non-frustrated random ferromagnet q=3 Potts model.
The {\it bicritical fixed point} ($N$) at
$(x^\star,T^\star)\sim(0.88,0.68)$, 
$(r^\star,f^\star)=(1.01(3),0.43(3))$
is characterized by the eigenvalues 
$\lambda_1=0.55-0.65,\ \lambda_2=0.30-0.40$ using the lattice sizes 
$L=6$ and $8$. However, at this fixed point a standard scaling plot
of $f$ approximately along the eigendirection of $\lambda_1$ and including
$L=10$ yields $1/\nu\sim 0.75-0.8$. Hence, we estimate
$\nu=1.28-1.36$, and using $[<m^2>]$ we find $\eta=0.17-0.22$. 
Both of these exponents are remarkable similar to the ones determined at the
bicritical point in the 2D random Ising model~\cite{singh} of
$\nu=1.32(8), \eta=0.20$. 
The {\it zero temperature fixed point} ($Z$) located at 
$x^\star\sim0.88(1)$, 
$r^\star\sim 0.96$ and we determine the exponents
$\nu=1.45-1.55, \eta=0.18(1)$, where we for the Ising model 
find $x^\star=0.89(1)$, $\nu=1.35-1.45,
\eta=0.18(1)$, in agreement with McMillan~\cite{McMill}. 
At this fixed point we also estimate $\theta \sim 0.3$. Since this exponent is
{\it positive}, temperature is irrelevant at this fixed point.

Several conclusions can now be drawn for the first time
for the randomly frustrated 3-state Potts model from
the results in
Fig.~\ref{fig:q3diag}.
The random critical fixed point $S$ has exponents 
that to within our precision appear
close to those of the pure Ising model, and in good agreement with theoretical
predictions~\cite{ludwig,dotsenko}.  
However, we find $\sigma(\Delta F) \ne 0$ at $S$, 
which certainly distinguishes it
from the pure Ising critical fixed point. At $S$ we have
also determined the fraction of negative 
$\Delta F$ 
for the fixed point
distribution. This fraction,  which is a measure of the ``renormalized''
frustration level, is decreasing with increasing $L$ at $S$ indicating that
the fixed point is the same as for unfrustrated ferromagnetic disorder.
A new bicritical fixed point $N$ appears even though the model is not
gauge-symmetric~\cite{nishimori}. This fixed point has exponents 
similar to the bicritical fixed point for the Ising model~\cite{singh}.
At the zero-temperature fixed point $Z$ temperature is irrelevant as it is
for the Ising model. The critical concentration $x_c$ is at this
point identical for the Ising and $q=3$ 
Potts 
model to within the numerical
resolution. Recently, Nishimori~\cite{N2} has rigorously established 
that $x_c^{\rm non-Ising} \ge x_c^{\rm Ising}$ for any bimodal vector-spin model
that is invariant under the operation ${\bf S_i}\to {\bf -S_i}$.
However, the $q=3$ 
Potts 
model is {\it not} invariant under this operation
and it is therefore remarkable that we for the $q=3$ model find 
that the inequality
apparently is satisfied as an equality. 
Finally, it would appear that the
$q=3$ 
Potts 
model is slightly reentrant in Fig.~\ref{fig:q3diag}, as 
occurs for the same model on hierarchical lattices~\cite{us}. However,
here, we believe that this is due to the fact that corrections to scaling
are more pronounced at low temperatures, in particular for a bimodal
distribution. It seems also possible that the
phase diagram could show reentrance in $(r,f)$ space but not in 
$(x,T)$ space.
For the bimodal Ising model it
is known that the phase boundary is vertical~\cite{N3} down from the bicritical point
($N$) where 
the Nishimori line~\cite{nishimori} intercepts the phase boundary. 
Remarkably, 
this is also what we find in the $(x,T)$ space for the $q=3$ Potts model.
Summarizing, our results indicate that at $N$ and $Z$
the $q=2$ (Ising) and $q=3$ {\it random} bimodal 
Potts 
models
are remarkably similar. However, for ferromagnetic disorder the Ising model
does not have a stable random fixed point at finite
temperature~\cite{ludwig,dotsenko}. 
The stable random fixed point
($S$) of the $q=3$ model has critical exponents very close
to the pure $q=2$ Ising model.
In light
of these observations we
believe that the fixed point structure shown in Fig.~\ref{fig:q3diag} is
rather general for two-dimensional disordered models when disorder is relevant.

We thank S.~M.~Girvin,
M.~P.~A.~Fisher and A.~P.~Young for helpful discussions.
This work has been supported in part by NSF grant number NSF DMR-9416906
and the NSERC of Canada.


\begin{references}

\bibitem{SGs}K. Binder and A.P. Young, Rev. Mod. Phys. {\bf 58}, 801
(1986); K.H. Fischer and J.A. Hertz, {\it Spin Glasses},
(Cambridge University Press, Cambridge,  1991).


\bibitem{univ}
For a discussion see J.~Cardy, J.\ Phys.\ A {\bf 29}, 1897 (1996); J.~Cardy and J.~L.~Jacobsen,
Phys.\ Rev.\ Lett. {\bf 79}, 4063 (1997); and references therein.


\bibitem{LFRG}
M.~P.~A.~Fisher, P.~B.~Weichman, G.~Grinstein and D.~S.~Fisher, Phys.\ Rev.\ B
{\bf 40}, 546 (1989);
D.~H.~Lee, Z.~Wang, and S.~Kivelson, Phys.\ Rev.\
Lett. {\bf 70}, 4130 (1993);
A.~W.~W.~Ludwig, M.~P.~A.~Fisher, R.~Shankar, and G.~Grinstein, Phys.\ Rev.\
B {\bf 50}, 7526 (1994). 



\bibitem{ludwig}
A.~W.~W.~Ludwig, Nucl.\ Phys.\ B {\bf 285}[FS19], 97 (1987);
A.~W.~W.~Ludwig, Phys.\ Rev.\ Lett.\ {\bf 61}, 2388 (1988).



\bibitem{shankar}
B.~N.~Shalaev, Sov.~Phys.~Solid State {\bf 26}, 1811 (1984);
R.~Shankar, Phys.~Rev.\ Lett. {\bf 58}, 2466 (1987);
{\it ibid.} {\bf 61}, 2390 (1988); 



\bibitem{dotsenko}
Vik.~S.~Dotsenko and Vl.~S.~Dotsenko, Adv.\ Phys.\ {\bf 32}, 129 (1983);
Vl.~S.~Dotsenko, M.~Picco, P.~Pujol, Phys.\ Lett.\ B {\bf 347}, 113 (1995);
Nucl.\ Phys.\ B {\bf 455}, 701 (1995).


\bibitem{luther}
G.~Grinstein and A.~Luther, Phys.\ Rev. B {\bf 13}, 1329 (1976).



\bibitem{singh}
R.~R.~P.~Singh, Phys.\ Rev.\ Lett.\ {\bf 67},899 (1991); R.~R.~P.~Singh
and J.~Adler, Phys. Rev. B {\bf 54}, 364 (1996).



\bibitem{McMill}
W.~L.~McMillan, Phys.\ Rev.\ B {\bf 29}, 4026 (1984);
{\it ibid.} {\bf 30}, 476 (1984).



\bibitem{nishimori}
H. Nishimori, J. Phys. Soc. Jpn. {\bf 55}, 3305 (1986);
H. Nishimori, J. Phys. Soc. Jpn. {\bf 61}, 1011 (1992);
P. Le Doussal and A. Georges, (unpublished);
A. Georges, D. Hansel, P. Le Doussal, and J.P.
Bouchaud, J. Phys. (Paris) {\bf 46} 1827 (1985).



\bibitem{harris}
A.~B.~Harris, J.\ Phys.\ C {\bf 7}, 1671 (1974).

\bibitem{Binder_Reger}K. Binder and J. Reger, Adv. Phys. {\bf 41}, 547 (1992).

\bibitem{denNijs}
H.~Park and M.~den~Nijs, Phys.\ Rev.\ B {\bf 38}, 565 (1988).

\bibitem{TM-refs}
I. Morgenstern and K. Binder, Phys. Rev. B {\bf 22}, 288 (1980);
H.~Saleur and B.~Derrida, J.\ Phys.\ (Paris) {\bf 46} 1043 (1985).

\bibitem{Night}
P.~Nightingale, J.\ Appl.\ Phys.\ {\bf 53}, 7927 (1982).

\bibitem{Cardy}
J.~Cardy, J.\ Phys.\ A {\bf 17} L385 (1984); Nucl.\ Phys.\ B {\bf
270}[FS16], 186 (1986); {\it ibid} {\bf 275}[FS17], 200 (1986).



\bibitem{Abraham}
D.~B.~Abraham and N.~M.~Svrakic, Phys.\ Rev.\ Lett. {\bf 56}, 1172
(1986).



\bibitem{Mon}
K.~K.~Mon and D.~Jasnow, Phys.\ Rev.\ A {\bf 30}, 670 (1984);
{\it ibid} {\bf 31}, 4008 (1985); D.~Jasnow, K.~K.~Mon, M.~Ferer,
Phys.\ Rev.\ B {\bf 33}, 3349 (1986).



\bibitem{FisherHuse}
D.~S.~Fisher and D.~A.~Huse, Phys.\ Rev.\ Lett.\ {\bf 56}, 1601 (1986);
Phys.\ Rev.\ B {\bf 38}, 386 (1988).

\bibitem{BrayMoore}
A.~J.~Bray and M.~A.~Moore, in {Heidelberg Colloqium on Glassy
Dynamics}, edited by  J.~L.~van Hemmen and I.~Morgenstern (Springer,
Berlin, 1987).

\bibitem{Pico}
M.~Picco, Phys. Rev. Lett. {\bf 79}, 2998 (1997).


\bibitem{N2}
H.~Nishimori, J.\ Phys.\ Soc.\ Jpn.\ {\bf 61}, 1011 (1992).


\bibitem{us}
M.~J.~P.~Gingras and E.~S.~S\o rensen, to be published.

\bibitem{N3}
Y.~Ozeki and H.~Nishimori, J.\ Phys.\ A {\bf 26}, 3399 (1993);
H.~Kitani, J.\ Phys.\ Soc.\ Jpn.\ {\bf 61}, 4049 (1992).

\end{references}
\end{document}